\documentclass[twocolumn,aps,pra,showpacs,showkeys,preprintnumbers,amsmath,amssymb]{revtex4}
\usepackage{graphicx}
\usepackage{dcolumn}
\usepackage{bm}
\usepackage{epsfig,wrapfig}
\usepackage{epstopdf} 
\usepackage{fancyhdr}
\usepackage{indentfirst}
\usepackage{titlesec}
\usepackage{amsmath}
\begin{document}
\title{Seismic Negative Belt of Acoustic Metamaterials}
\author{Sang-Hoon  \surname{Kim}$^{a}$} \email{shkim@mmu.ac.kr }
\author{Mukunda P. \surname{Das}$^{b}$}\email{mukunda.das@anu.edu.au}
\affiliation{
$^a$Division of Marine Engineering, Mokpo National Maritime University,
Mokpo 58628, R. O. Korea
\\
$^b$Department of Theoretical Physics, RSPE,
The Australian National University, Canberra, ACT 0200, Australia
}
\date{\today}
\begin{abstract}
An earthquake-proof seismic negative belt of an artificial seismic shadow zone is introduced.
The belt is composed of acoustic materials
which has one of the constituent parameter between density and modulus is negative effectively.
It converts the velocity of the seismic wave imaginary,
and then creates a stop-band for the seismic frequency range.
The belt is an attenuator of a seismic wave that reduces the amplitude of the wave exponentially.
Passing the belt underground, the seismic energy turns into sound and heat in air
and the magnitude of the seismic wave is weakened to be defended by conventional method.
Models of the negative density and negative modulus for engineering are suggested.
\end{abstract}
\date{\today}
\maketitle\

In human history there have been three natural disasters:
Asteroid collisions, Earthquakes including tsunamis, and biological epidemics.
Asteroid collision is very rare during one human life cycle
and most of biological epidemics are under control nowadays.
However, in spite of great developments of earthquake engineering,
earthquake is still a serious threat to human civilization and leaves several thousand deaths every year.

Earthquakes are the result of sudden release of huge amount of energy in the Earth's crust that produces seismic waves.
The hypocenter or focus is the point where the stored strain energy is first released
and the earthquake rupture begins.
Most hypocenters are located at the very upper parts of the mantle of the Earth.
 A seismic wave is a kind of acoustic wave and there are two types of seismic waves:
 body waves and surface waves\cite{vill}.
 Primary(P) and Secondary(S) waves are body waves, and  Rayleigh(R) and Love(L) waves are surface waves.
The surface wave is generated when the body waves arrive at the surface of the earth
and the epicenter is the main point of the generation.
The amplitudes of the surface waves decrease exponentially with the depth
but decays more slowly with the distance than that of body waves\cite{udias,aki}.

R waves can exist only in a homogeneous medium with a boundary and
  have transverse motion\cite{vill,udias,aki}.
 Earthquake motions observed at the ground surface are mainly due to R waves.
L waves are polarized shear waves guided by an elastic layer.
It is this fact that causes horizontal shifting of the Earth during earthquakes.
L waves have both longitudinal and transverse motion and this is
what most people feel directly during earthquakes.
Surface waves travel  slowly  as $1 \sim 3 km/sec$,
but the wavelengths are large as of the order of $100 m$ compared with the body waves.
The frequencies of seismic waves are just below the audible frequency or infrasound.
Therefore, earthquake engineering is not to defend the vertical body waves but to defend the horizontal surface waves.
By these reasons epicenter is more important than hypocenter in earthquake engineering.

Conventional earthquake engineering is, in principle, the engineering of vibration and passive responses.
It is aimed to ensure that the buildings themselves do not collapse from swinging or vibrations.
These methods have more than thousand year history and are pretty effective,
 but still a lot of people die from earthquakes every year.
Recently totally new methods based on acoustic metamaterials have been introduced\cite{sebas1,sebas2,kim1,kim3}.
Metamaterials are man-made effectively homogeneous structures
 with dimensions potentially much smaller than  that of a wavelength.

There are two approaches in the metamaterial methods in earthquake engineering.
One is an acoustic cloaking method to protect an individual building like the previous conventional method\cite{sebas1,sebas2}.
It is omnidirectional and there is no energy dissipation by the impedance matching.
However, it is a point protection which are good for buildings smaller than seismic wavelength.
The other is an artificial seismic shadow zone method to protect a whole city instead of individual buildings\cite{kim1,kim3}.
It builds an acoustic attenuator underground by creating a stop-band of the seismic wave.
In this paper we focus the second method because it is the only macroscopic protection known till now\cite{sheng}.

Seismic wave is a kind of acoustic wave and its velocity is decided by the two constituent parameters of the medium:
the density $\rho$ and the compressibility $\kappa$ or the inverse of the modulus $E$ as
\begin{equation}
v= \frac{1}{\sqrt{\kappa \rho}} = \sqrt{\frac{E}{\rho}}\hspace{0.1 cm},
\label{10}
\end{equation}
where $\kappa$ and $\rho$ and the electro-optic analog of electric permittivity $\epsilon$ and the magnetic permeability $\mu$.
The modulus  $E$ could be any type:
Young's modulus at 1D, Shear modulus at 2D, and Bulk modulus at 3D.

If one of the acoustical constituent parameter is negative effectively at some specific frequency ranges,
then the velocity in Eq. (\ref{10}) becomes imaginary in the frequency range and it makes the wave decay exponentially.
Therefore, negative density or negative modulus creates the stop-band of the wave and the key of the seismic negative belt
that dissipates seismic energy.
Note that the acoustical impedance $Z= \rho v = \sqrt{\rho E}$ becomes imaginary because it is an energy absorption.
If the two constituent parameters are negative simultaneously, it creates a backward wave without the dissipation.

The effective negative density has been realized already\cite{milton,huang,oudich,fino}.
The concept of the negative density or negative inertia is pretty simple.
From the Newton's law, the mass $m$ and moment of inertia $I$ are related with force and torque as
\begin{equation}
-m=\frac{F}{-a}
\hspace{0.3 cm} {\rm and} \hspace{0.2 cm}
-I=\frac{\tau}{-\alpha} \hspace{0.1 cm},
\label{16}
\end{equation}
where $a$ and $\alpha$ are the linear and angular acceleration.
If the mass responds against to the applied force or torque,
it acts as a negative mass or negative moment of inertia effectively.
The negative density or inertia creates an anti-phase motion and decays of the wave.

\begin{figure}
\resizebox{!}{0.11\textheight}{\includegraphics{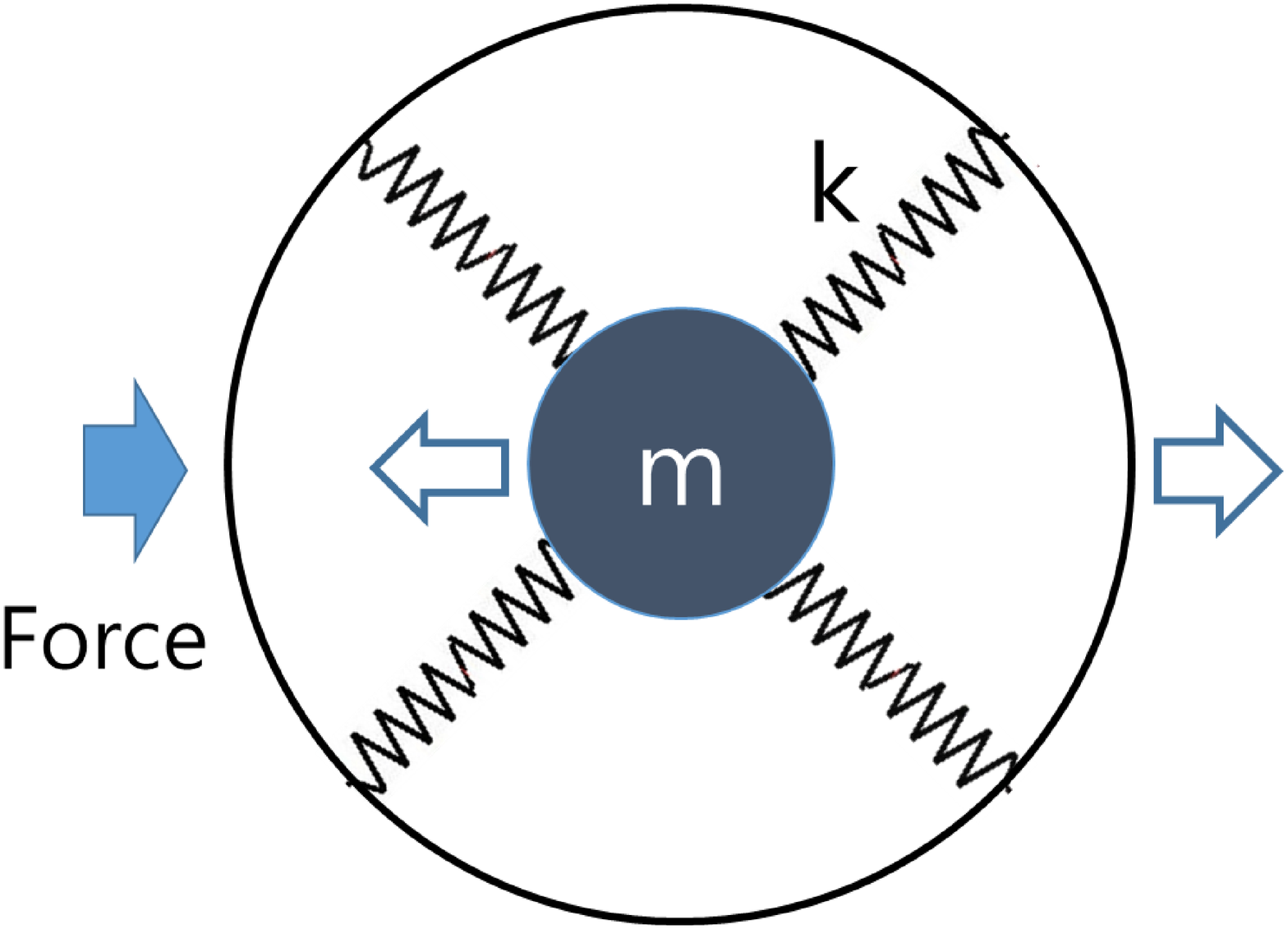}}
\resizebox{!}{0.15\textheight}{\includegraphics{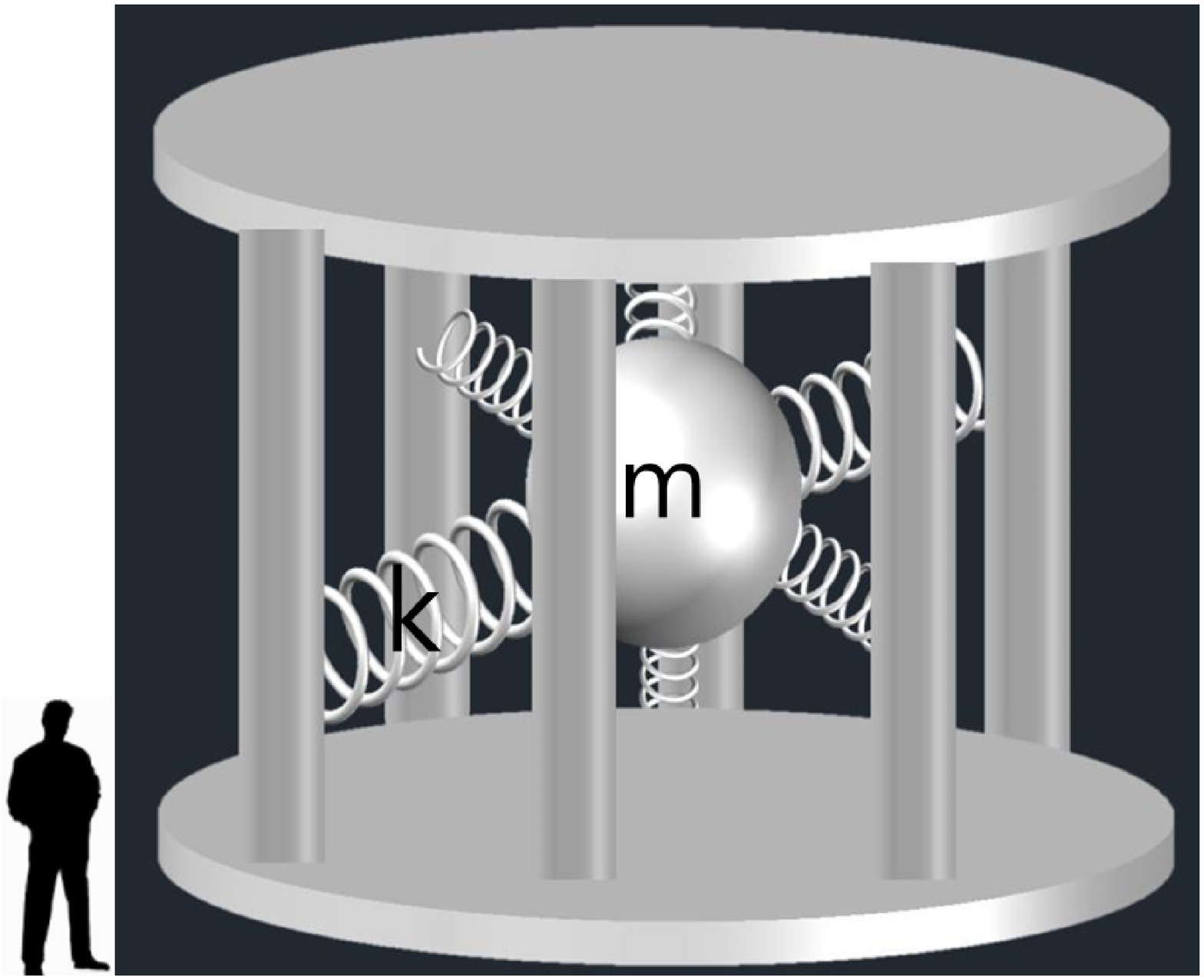}}
\caption{Concept of the negative density and its model of the negative ball.
The white arrows are the direction of the acceleration to the force from the left side.}
\label{density}
\end{figure}

An effective negative density or negative mass is a multiple mass or mass-in-mass system
connected by springs as in Fig.~\ref{density}.
If a force or torque is applied to the outer bone structure,
then the inertia of the ball inside reacts against the force or torque.
They oscillate against external force creating an anti-phase motion, and then cancel out the seismic vibration.
 A simple harmonic motion of a spring-mass system  which acts as $a=-\omega_o^2 x$
 is the simplest case of the negative density.
A membrane structure behaves an effective negative density, too.
The mass of the central ball in Fig.~\ref{density} is $10^2 \sim 10^3$ kg,
and the spring constants are $10^5 \sim 10^7$ N/m.
The oscillation frequency of the structure matches with the seismic surface waves: $\omega_o =\sqrt{k_{eff}/m}$,
where $k_{eff}$ is the effective spring constant.
The system should have different oscillation frequencies because the surface waves are inhomogeneous.

The application of the negative density is found everywhere.
It has been used as a damper for very high buildings.
The negative belt is to build the damper region underground between epicenters and residential areas.
There are enough data of epicenters for one hundred years,
 and new earthquakes take place around previous epicenter region again like volcanoes.
The underground negative belt composed of effective negative mass structures
dissipates seismic energy and weakens the earthquakes.

\begin{figure}
\resizebox{!}{0.2\textheight}{\includegraphics{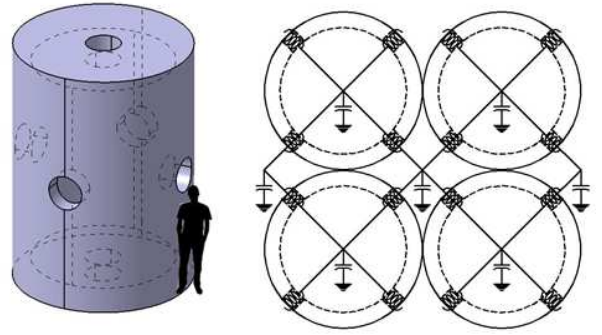}}
\caption{Model of meta-box which has negative modulus and its electrical analogy\cite{kim1}.
The empty spaces among the cylinders act as a large capacitor.}
\label{modulus}
\end{figure}

An effective negative modulus has been realized by mechanical resonances\cite{cheng2,wu2}.
The resonance of accumulated waves in the resonator reacts against the applied pressure at some specific frequency ranges.
Then, the negative modulus is realized by passing the acoustic wave through an array of Helmholtz resonators.
In this way the acoustic intensity decays just above the resonant frequencies\cite{kim1,kim3}.
Multi resonance frequencies creates multi stop-band of the wave.
The Helmholtz resonator is a realization of an electrical resonance circuit by mechanical correspondence.
A pipe or tube with open both ends corresponds to an inductor.
On the other hand, one open and one closed end corresponds to a capacitor.
The resonance frequency of a Helmholtz resonator is\cite{kim1,kim3}
\begin{equation}
\omega_o = c \sqrt{\frac{S}{l' V}} \hspace{0.1 cm}.
\label{20}
\end{equation}
$S$ is the area of the entrance,  $V$ is the volume, and $c$  is the background  velocity.
$l'$  is the effective length which is given by  $l' \simeq l + 0.85d$,
 where  $l$ is the length of the hole or thickness of the resonator and  $d$ is the diameter of the hole\cite{ever,bera}.

We designed a meta-box based on the Helmholtz resonator Fig.~\ref{modulus}
 to match the seismic surface frequencies\cite{kim1}.
 The size can be estimated by analogy between electric circuits and mechanical pipes.
 The diameter of the hole and the thickness of the cylinder is order of foot,
 and the volume inside is order of $10 \sim 100 m^3$.
Because the wavelength of the seismic wave is much longer than the size of the holes,
the incoming wave will be diffracted strongly into the meta-boxes.
Note that the more the number of holes, the higher the resonant frequencies.
The shape of the meta-box is neither necessary to be circular nor to have 6 holes.
It could be any form of a concrete box with a few side holes such as cubic or hexagonal shapes.

Resonances are common in everyday life.
Laser has the resonance of atoms, microwave oven has the resonance of molecules and the wind instrument has
 the resonance of artificial atoms.
The negative modulus method is something like to build hundred thousands of huge
concrete-made saxophones underground and play them with seismic energy.
The energy turns into sound and heat.
The negative modulus method has been used for long time in civil engineering.
There has occurred a large earthquake in Mexico in 1985,
and it was found that some buildings of specific size destroyed completely
compared with other buildings were relatively safe.
It was the seismic energy dissipation by resonances.
It is common sense for civil engineers to avoid resonances from external forces or torques including seismic waves.
The negative belt of the effective negative modulus is the reverse concept of it.

\begin{figure}
\resizebox{!}{0.23\textheight}{\includegraphics{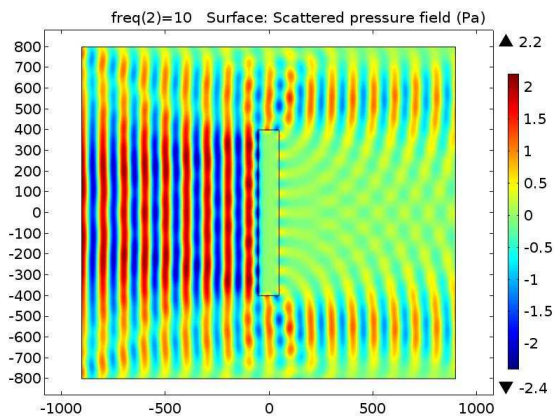}}
\caption{Pressure distribution by a negative belt.
Acoustic wave comes from the left side. Freq.$=10 Hz$. The units are m.}
\label{belt}
\end{figure}

We build a negative belt or an attenuator of the seismic surface wave
 by filling up many resonators or ``meta-boxes" underground.
 The amplitude of the seismic wave that passes through the negative belt is reduced as in Fig.~\ref{belt}.
Epicenters are located in the left side of the belt and the other side is the seismic shadow zone.
The energy of the seismic waves is dissipated inside of the meta-boxes,
and the absorbed energy is converted into sound and heat.
The use of a variety of resonators will cover a wide range of seismic wave frequencies.

A numerical simulation of the negative belt is shown in Fig.~\ref{belt}.
We applied some typical material parameters.
 Density and modulus are $\rho=1.0 \times 10^3 kg/m^3$, $E=1.0 GPa$ for the background.
It makes the incoming seismic wave velocity of $1 km/sec$.
The background could be rocks, water, or mixed.
We applied $\rho=-2.0 \times 10^3 kg/m^3$ and $E=-0.50 GPa$ for the negative belt,
and $E=0.50 GPa$ with the same density for the normal barrier.
Note that the relative refractive index is $2i$ for the negative belt.
The impedance of the belt is imaginary or absorption of energy.
It roles as an ``attenuator" to the seismic waves.
The left border of the belt in Fig.~\ref{belt} is breaking by the impedance difference,
but could be fixed after the earthquakes.

Besides the large scale of the protection, there is more advantage of the negative belt.
It is the construction cost at large magnitude of earthquakes.
The common form of the seismic magnitude, $M$, in {\it Richter-scale}
  defined by comparing the two amplitudes as
\begin{equation}
M=\log \frac{A}{A_o}\hspace{0.1 cm},
\label{32}
\end{equation}
where $A$ is the maximum amplitude of the seismic wave which is rescaled at $100 km$  from the epicenter.
 $A_o$ is the background amplitude to be set $A_o = 1 \mu m$.
 In principle the conventional earthquake engineering is to reduce the amplitude of buildings be seismic wave.
Then, the cost increase exponentially as the magnitude $M$ is increasing.
Magnitude 3 or lower earthquakes are almost imperceptible and rarely cause damage.
 On the other hand,  magnitude 6 or higher ones are very destructive,
and it is expensive to build.

\begin{figure}
\resizebox{!}{0.23\textheight}{\includegraphics{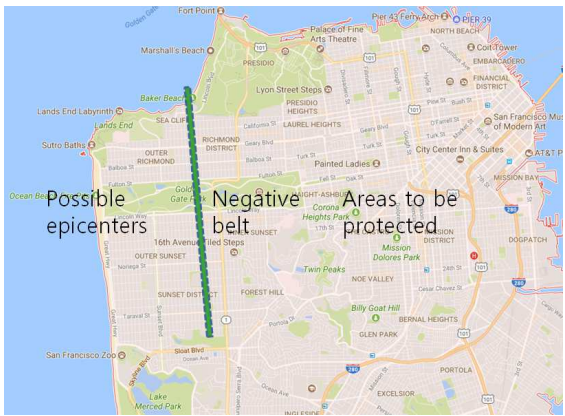}}
\caption{Possible position of the negative belt to protect San Francisco, U. S. A.
The epicenters are located in the Ring of Fire.}
\label{sanf}
\end{figure}

Let the initial seismic wave, that is, before entering the waveguide, have
amplitude $A_i$ and magnitude $M_i$,
and final seismic wave, that is after leaving the waveguide,
 have amplitude $A_f$ and magnitude $M_f$ following the Eq. (\ref{32}).
If we assume that the plain seismic wave of wavelength $\lambda$
propagates in $x-$direction, the amplitude of the wave reduces exponentially as
\begin{equation}
 A_f = A_i  e^{-2\pi |n| x/\lambda}\hspace{0.1 cm},
 \label{20}
\end{equation}
where and $\lambda$ is the wavelength of the seismic surface wave,
 and $n$  is the relative refractive index of the negative belt.
The amplitude of the seismic wave reduces exponentially as passing the waveguide
of metamaterials.
We can rewrite Eq. (\ref{20}) with the definition of
the magnitude in Eq. (\ref{32}) as
\begin{equation}
A_o 10^{M_f} = A_o 10^{M_i}  e^{-2\pi |n| x/\lambda}\hspace{0.1 cm},
\label{30}
\end{equation}
where $W$ is the width of the belt or $x$.
Taking logarithms both sides of Eq. (\ref{30}), we obtain the width  as
\begin{equation}
W = \frac{0.366 \lambda }{|n|}\Delta M\hspace{0.1 cm},
\label{34}
\end{equation}
where $\Delta M = M_i - M_f$.
For example, if $n = 1.5$ and $\lambda = 100 m$, then $W = 25 m$ for $\Delta M = 1$.
Therefore, the cost of the negative belt is linearly increasing as the magnitude is increasing.
$\Delta M = 4$  can be engineered by at most  $W = 100 m$ in Eq. \ref{34}.
If seismic wave of magnitude 8 comes in the belt, then it leaves the belt with magnitude 4.
The magnitude 4 is not serious and can be defendable by conventional methods of earthquake engineering.
The depth of the belt should be at least the same as the foundation of the building,
 but it is not necessary to be more than the wavelength of the seismic surface waves.
 Note that a high refractive index (soft but heavy) material is desirable for the narrow seismic barrier.
Composite materials could be good candidate for this belt.
We suggested a negative belt to protect residential area in San Francisco from earthquakes in Fig.~\ref{sanf}.
The epicenters are a part of the Ring of Fire and located in the left side of the belt.

A negative belt method of acoustic metamaterials for the attenuation of seismic waves are introduced.
This method which is different from the conventional cloaking technology produces an artificial seismic shadow zone that reduces the seismic wave using effective negative density or modulus.
Some engineering models of negative density and negative modulus were shown.
The negative belt is constricted underground anyplace between possible epicenters and the areas to be protected.
 It changes the seismic wave-vector to an imaginary one and reduces the amplitude of the seismic wave exponentially.

 A numerical simulation of the seismic shadow zone was conducted
 and a possible application to protect San Francisco in U. S. A. was suggested.
The seismic range of the structure can be upgraded or downgraded by adjusting the width of the belt.
The biggest advantage is that all structures behind the barrier be protected, which decreases the cost of seismic protection.
We hope to test this method with engineers using small scale test models soon.

Tsunamis are commonly accompanied by  earthquakes and causes damage far greater than earthquakes.
 Therefore, we can not separate tsunami from earthquakes.
 Recently, we have published a totally different tsunami wall method by meta-lens in the sea\cite{kim5}.
 This is a removable tsunami wall that is based on an optical Eaton lens.
The principle is to transform the incoming wave into a rotating wave
and the impedance matching is used to reduce the pressure of the tsunami without reflecting the wave.
It is just a beginning and  will be a next hot subject in acoustic metamaterials beyond earthquake.


\end{document}